\documentclass[iop,apj,tighten,numberedappendix]{emulateapj}
\usepackage[breaklinks,colorlinks,urlcolor=blue,citecolor=blue,linkcolor=blue]{hyperref}

\usepackage{amsmath,amssymb}
\usepackage{cleveref}
\usepackage{graphicx}
\usepackage{bm}
\usepackage[toc,title,page]{appendix}
\usepackage{tocbibind}
\usepackage{color}
\newcommand{\Msun}{{\rm M}_\odot}

\def\lsim{~\rlap{$<$}{\lower 1.0ex\hbox{$\sim$}}}
\def\gsim{~\rlap{$>$}{\lower 1.0ex\hbox{$\sim$}}}

\shorttitle{Signature from BHs in AGNs}
\shortauthors{Tagawa~et~al.}

\begin{document}
\title{
Observable signatures of stellar-mass black holes in active galactic nuclei
}

\author{
Hiromichi Tagawa\altaffilmark{1,2}, 
Shigeo S Kimura\altaffilmark{2,3},
Zolt\'an Haiman\altaffilmark{1,4},
Rosalba Perna\altaffilmark{5,6},
Imre Bartos\altaffilmark{7}
}
\affil{
\altaffilmark{1}Department of Astronomy, Columbia University, 550 W. 120th St., New York, NY, 10027, USA\\
\altaffilmark{2}Astronomical Institute, Graduate School of Science, Tohoku University, Aoba, Sendai 980-8578, Japan\\
\altaffilmark{3}Frontier Research Institute for Interdisciplinary Sciences, Tohoku University, Sendai 980-8578, Japan\\
\altaffilmark{4}Department of Physics, Columbia University, 550 W. 120th St., New York, NY, 10027, USA\\
\altaffilmark{5}Department of Physics and Astronomy, Stony Brook University, Stony Brook, NY 11794-3800, USA\\
\altaffilmark{6}Center for Computational Astrophysics, Flatiron Institute, New York, NY 10010, USA\\
\altaffilmark{7}{Department of Physics, University of Florida, PO Box 118440, Gainesville, FL 32611, USA}\\
}
\email{E-mail: htagawa@astr.tohoku.ac.jp}

\begin{abstract} 
Stellar-mass black holes (BHs) are predicted to be embedded in the disks of active galactic nuclei 
(AGN) due to gravitational drag and in-situ star formation. 
However, clear evidence for AGN disk-embedded BHs is currently lacking.
Here, as possible electromagnetic signatures 
of these BHs, we investigate breakout emission from shocks emerging around Blandford-Znajek jets
launched from accreting 
BHs in AGN disks. 
We assume that the majority of the highly super-Eddington flow reaches the BH, 
produces a strong jet, and the jet produces feedback that shuts off accretion and thus leads to episodic flaring. 
While these assumptions are highly uncertain at present, they predict a breakout emission 
characterized by luminous
thermal emission in the X-ray bands, and bright, broadband non-thermal emission from the infrared to the gamma-ray bands. 
The flare
duration depends on the BH's distance $r$ from the central supermassive BH, varying between $10^3-10^6$~s for  $r \sim 0.01-1$~pc.
This emission can be discovered by current and future infrared, optical, and X-ray wide-field surveys and monitoring campaigns of nearby AGNs. 
\end{abstract}

\section{Introduction}

It is a common belief that stars and compact objects (COs), including stellar-mass black holes (BHs), are embedded in the disks of active galactic nuclei (AGNs) due to capture via dynamical interactions between the nuclear star cluster (NSC) 
and the AGN disk \citep{Ostriker1983,Syer1991}, and in-situ star formation \citep{Levin2003,Goodman04,Thompson05,Levin2007}. 
There are several observations supporting this picture. 
The high metallicity of quasars is presumably related to frequent explosive phenomena of COs and stars in AGN disks \citep{Artymowicz1993,Wang2010,Xu18,Wang2021_Metallicity,Toyouchi22_AGN_IMF}. 
The existence of young stars \citep{Genzel2003,Levin2003} and clusters \citep{Milosavljevic2004} around Sgr A*, as well as the 
high metallicity component of NSCs \citep{Antonini2015,Do20,Neumayer20,Fahrion2021}
 imply that 
stars, and hence COs, 
form in-situ in AGN disks. 
Furthermore, the spatial distribution of low-mass X-ray binaries discovered in the Galactic center \citep{Hailey18,Mori2021} is consistent with the evolution of COs and stars in an AGN disk \citep{Tagawa19}.

AGN disks are plausible environments for BH-BH \citep[e.g.][]{Bartos17,Stone17,McKernan17,Yang19b_PRL,Tagawa19} 
and BH-neutron star (NS) mergers \citep{McKernan2020_BHNSWD,Tagawa20_MassGap,Yang20_GW190814} reported as gravitational wave (GW) events by the LIGO \citep{2015CQGra..32g4001L}, Virgo \citep{2015CQGra..32b4001A} and KAGRA \citep{KAGRA} detectors \citep{Venumadhav19,LIGO20_O3_Catalog,Abbott21_GWTC3}. 
This pathway can explain the distributions of masses, mass ratios \citep{Yang20_gap,Gayathri2021_AGN_O3}, 
spin vectors \citep{Tagawa20b_spin}, 
and correlation between the masses and spin magnitudes \citep{Tagawa2021_hierarchical}
for the bulk of merging events. 
Furthermore, AGN disks are promising environments to explain the characteristic properties, 
high mass \citep{Tagawa20_MassGap}, 
possible high eccentricity \citep[][]{Samsing20,Tagawa20_ecc,Romero-Shaw20,Gayathri2022}, 
and hypothesized electromagnetic (EM) counterpart, ZTF19abanrhr \citep{Graham20}, 
of the unexpected GW event GW190521 \citep[][]{LIGO20_GW190521}. 
In addition, the first GW event, GW150914 \citep{Abbott16a}, might be associated with a bright gamma-ray event, GW150914-GBM (\citealt{Connaughton2016,Connaughton2018}, 
but see \citealt[][]{Greiner2016,Xiong2016,Savchenko2016}), which may imply a merger in a gas-rich environment.

Recently, several studies have investigated emission from transients emerging from AGN disks. 
\citet{Zhu2021_Cocoon_NSMs}, \citet{Zhu2021_Neutrino}, 
\citet{Perna2021_GRBs}, 
\citet{Yuan2021}, 
\citet{WangPerna2022_GRBA}, 
and \citet{Lazzati2022} 
estimated the emission from gamma-ray bursts, and 
\citet{Perna2021_AICs} and 
\citet{Zhu2021_WD_AIC} discussed the electromagnetic signatures expected from accretion induced collapse of neutron stars and white dwarfs.
\citet{Yang2021_TDE} studied the properties of tidal disruption of stars by stellar-mass BHs, 
while
\citet{Grishin2021} investigated supernova explosions, 
and \citet{Bartos17} and \citet{Stone17} estimated the electromagnetic emission produced by thermal radiation and/or outflows from circum-BH disks in AGN disks. 
There are several studies which investigated possible transients from merging BHs in AGN disks, focusing on the association of the optical flare, ZTF19abanrhr, with the BH merger. 
\citet{McKernan2019_EM} discussed emission from shocks caused by 
collision between gas bound to the merged remnant and unbound gas after recoil kicks due to anisotropic radiation of GWs. 
\citet{Graham20} assessed the net luminosity and timescales for gas accretion induced by recoil kicks.
{
\citet{deMink2017} considered flares emerging from shocks in a circum-BH disk due to recoil kicks. 
}
\citet{Kimura2021_BubblesBHMs} 
and \citet{Wang2021_TZW,Wang2021b}, respectively, considered thermal and non-thermal emission from bubbles around BHs due to strong outflows 
considering continuous and episodic 
super-Eddington accretion, 
and 
\citet{Wang2021_TZW} further considered emission from shocks emerging due to interactions of Blandford-Znajek (BZ) jets \citep{Blandford1977} launched from accreting BHs to the broad line regions. 
\citet{Tagawa2022_BHFeedback} (hereafter Paper~I) estimated the structure of the cavity created by the BZ jet and dynamical evolution of gas around the BHs. 
\citet{Tagawa2023} (hereafter Paper~II) 
investigated the properties of emission from shocks emerging around jets launched from a BH merger remnant.

\begin{figure*}
\begin{center}
\includegraphics[width=145mm]{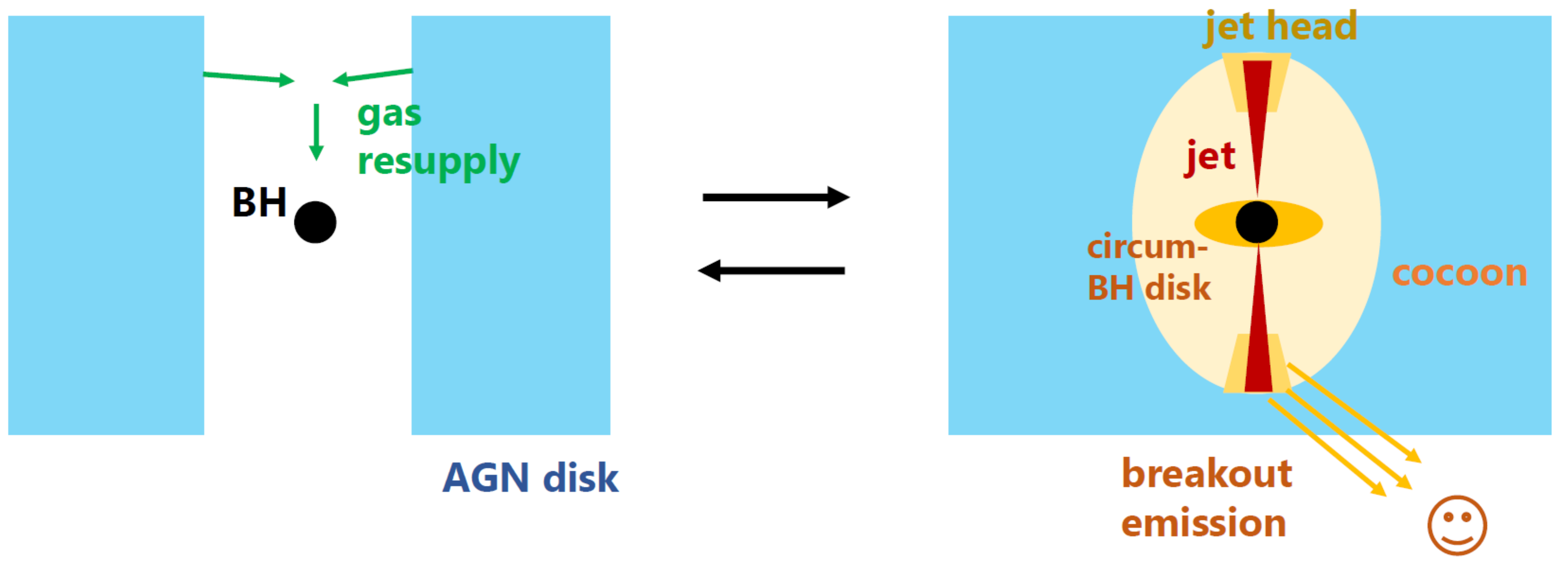}
\caption{
Schematic picture of quiescent phases (left) and  
the breakout emission from the head of a jet launched from a solitary BH (right) embedded in an AGN disk. 
}
\label{fig:schematic_solitary}
\end{center}
\end{figure*}

In this paper, we apply the method developed in Paper~II, and 
evaluate properties and observabilities of 
thermal and non-thermal emission from shocks emerging around jets launched by accreting solitary BHs due to the BZ effect (Fig.~\ref{fig:schematic_solitary}). 
We find that thermal emission is bright in X-ray bands, 
while non-thermal emission is bright in infrared to gamma-ray bands. 
This emission is predicted to be discoverable by current and future optical and X-ray telescopes, that is the Zwicky Transient Facility (ZTF), Vera Rubin, XMM-Newton, HiZ-GUNDAM, Einstein Probe, NuSTAR, FORCE, XRT, Chandra, JWST, and WISE.

\section{Emission}

In this section we describe the method for calculating the properties of the breakout emission produced from solitary BHs in AGN disks. More details
on this computation are provided in Paper~II.

\subsection{Mechanisms for breakout emission}
\label{sec:mechanisms}

Here we highlight the physical mechanisms responsible for producing the breakout emission from solitary BHs in AGN disks (see Fig.~\ref{fig:schematic_solitary} for a schematic representation). 
In these disks, isolated BHs are surrounded by circum-BH disks, since the gas captured by BHs from the AGN disk has enough angular momentum to circularize around the BHs \citep{Tanigawa2012}. 
When the circum-BH disk is advection dominated, as expected here, a magnetically dominated state can be realized \citep[e.g.][]{Meier2001,Kimura2021_BBH_PeV} owing to the accumulation of the magnetic flux in the vicinity of the BH \citep{Cao2011}. Even if the magnetic flux is initially weak, the outflow from the disk converts the toroidal magnetic field generated by the shear motion into a poloidal field \citep{Liska2020}. 
In these cases, jets from  spinning BHs can be launched through the BZ process \citep{Blandford1977}.
The jet power ($L_{\rm j}$) is proportional to the mass accretion rate onto the BH (${\dot m}$), \begin{align}L_{\rm j}=\eta_{\rm j}{\dot m} c^2,\end{align} where $\eta_{\rm j}$ is the conversion efficiency from rest mass to jet power, which is approximated by $\eta_{\rm j}\sim a_{\rm BH}^2$ for a magnetically dominated jet \citep[e.g.][]{Tchekhovskoy2010,Narayan2021}, $a_{\rm BH}$ is the dimensionless spin of the BH (see \S\,\ref{sec:numerical_choice_m6} for its choice), and $c$ is the speed of light. Since the power of a shock emerging around the jet and the luminosity of radiation emitted from the shock are roughly proportional to the jet power, the accretion rate onto the BH is a key quantity to determine the observed luminosity from the system.

The accretion rate onto a circum-BH disk in the AGN disk is often evaluated via a modified Bondi-Hoyle-Lyttleton (BHL) rate, as given by Eq.~(1) of Paper~I. To consider a possible reduction from the BHL rate, we parameterized the fraction of the accretion rate onto the BH (${\dot m}$) over the Bondi-Hoyle-Lyttleton rate (${\dot m}_{\rm BHL}$) as $f_{\rm acc}={\dot m}/{\dot m}_{\rm BHL}$. 
For example, low $f_{\rm acc}$ may be predicted due to winds from 
an accretion disk with a super-Eddington rate, although 
recent simulations suggest that the conversion to wind 
is moderate \citep{Kitaki2021} for accretion flows in which the circularization 
radius (where gas is circularized after being captured by a BH) is much larger than the trapping radius (within which photons are advected to a BH without escaping), as is the case for BHs embedded in an AGN disk.
In addition, 
the accretion rate onto a BH in a cavity during the
active phases is estimated to be lower by a factor of a few 
compared to that without a cavity \citep{Tagawa2022_BHFeedback}. 
As a fiducial value, we simply adopt $f_{\rm acc}=1$.

Once the jet collides with the AGN gas, 
a cocoon of shocked gas forms around the jet. 
Due to the high pressure of the cocoon, 
AGN gas around the BH, together with the outer regions of the circum-BH disk, are quickly evacuated. The BH keeps accreting and the jet remains active until the inner remnant regions of the truncated circum-BH disk are consumed by the accretion. Subsequently, the BH is quiescent and the cavity begins to fill in gradually. 
Finally, AGN gas is recaptured by the BH, and the cocoon reopens a cavity. 
We predicted in Paper~I that such a cycle repeats many times until the dissipation of the AGN disk.

As the jet collides with unshocked gas in the AGN disk, strong shocks form. 
During the early phases, photons in the shocked medium cannot
escape from the system because they are surrounded by
the optically thick AGN disk. 
As the shock approaches the surface of the AGN disk, 
thermal photons in the shocks begin escaping from the system, 
and non-thermal electrons begin to be accelerated due to the formation of collisionless shocks, leading to luminous thermal and non-thermal emission. 
As non-thermal emission, we take into account synchrotron radiation, synchrotron-self Compton scattering, and second-order inverse Compton scattering. Because of the high density of AGN gas, we need to consider synchrotron self-absorption.

In Paper~II we predicted the properties of the breakout emission emerging from merger remnant BHs, 
and the same formulae can be applied to the emission from solitary BHs. Hence here,
by applying the models constructed in Paper~II, 
we discuss the properties and the observability of the breakout emission from solitary BHs.

\subsection{Numerical choices}
\label{sec:numerical_choice_m6}

In the fiducial model
we adopt the same parameter values as in Paper~I. More specifically: 
the BH mass is $m=10\,\Msun$, 
the radial distance of the BH from the central SMBH is $R_{\rm BH}=1\,{\rm pc}$,
the mass of the SMBH is $M=10^6\,\Msun$, 
the gas inflow rate from the outer boundary ($R_{\rm out}=5\,{\rm pc}$) of the AGN disk 
is ${\dot M}_{\rm in}=1~L_{\rm Edd}/c^2$, 
where $L_{\rm Edd}$ is the Eddington luminosity of the SMBH, 
the angular momentum transfer parameter 
in the outer 
AGN disk is $m_{\rm AM}=0.15$ \citep{Thompson05}, 
the viscous parameter 
in the inner
disk is $\alpha_{\rm AGN}=0.1$ \citep{King07,Martin2019}, 
and the opening angle of the injected jet is $\theta_0=0.2$ \citep[e.g.][]{Pushkarev2009,Hada2013,Hada2018,Berger2014}. 

We set 
the jet energy conversion efficiency to $\eta_{\rm j}=0.1$
considering that spin-up by accretion and spin-down by the BZ jet may be roughly equal at around $a_{\rm BH}\lesssim 0.3$\footnote{
The spin magnitude of the BHs in the AGN disk is assumed to be lower than that observed for typical X-ray binaries \citep{Reynolds2021}. 
This is because the former keeps powering a BZ jet while accreting, while the other does not in soft states. The fiducial choice is conservative, and the breakout emission becomes brighter if the spin magnitude is higher. 
}
(e.g., Fig.~10 of \citealt{Narayan2021}), 
the fraction of postshock energy carried by the post shock magnetic field 
and by electrons 
to $\epsilon_{\rm B}=0.03$ (e.g., \citealt[][]{Panaitescu2001,Uchiyama2007,Santana2014}) and $\epsilon_{\rm e}=0.1$ (e.g., \citealt[][]{Waxman1999,Panaitescu2001,Sironi_2013,Santana2014}),
respectively, 
and the power-law slope for injected electrons accelerated by the first-order Fermi process to $p=2.5$.

\begin{figure}
\begin{center}
\includegraphics[width=75mm]{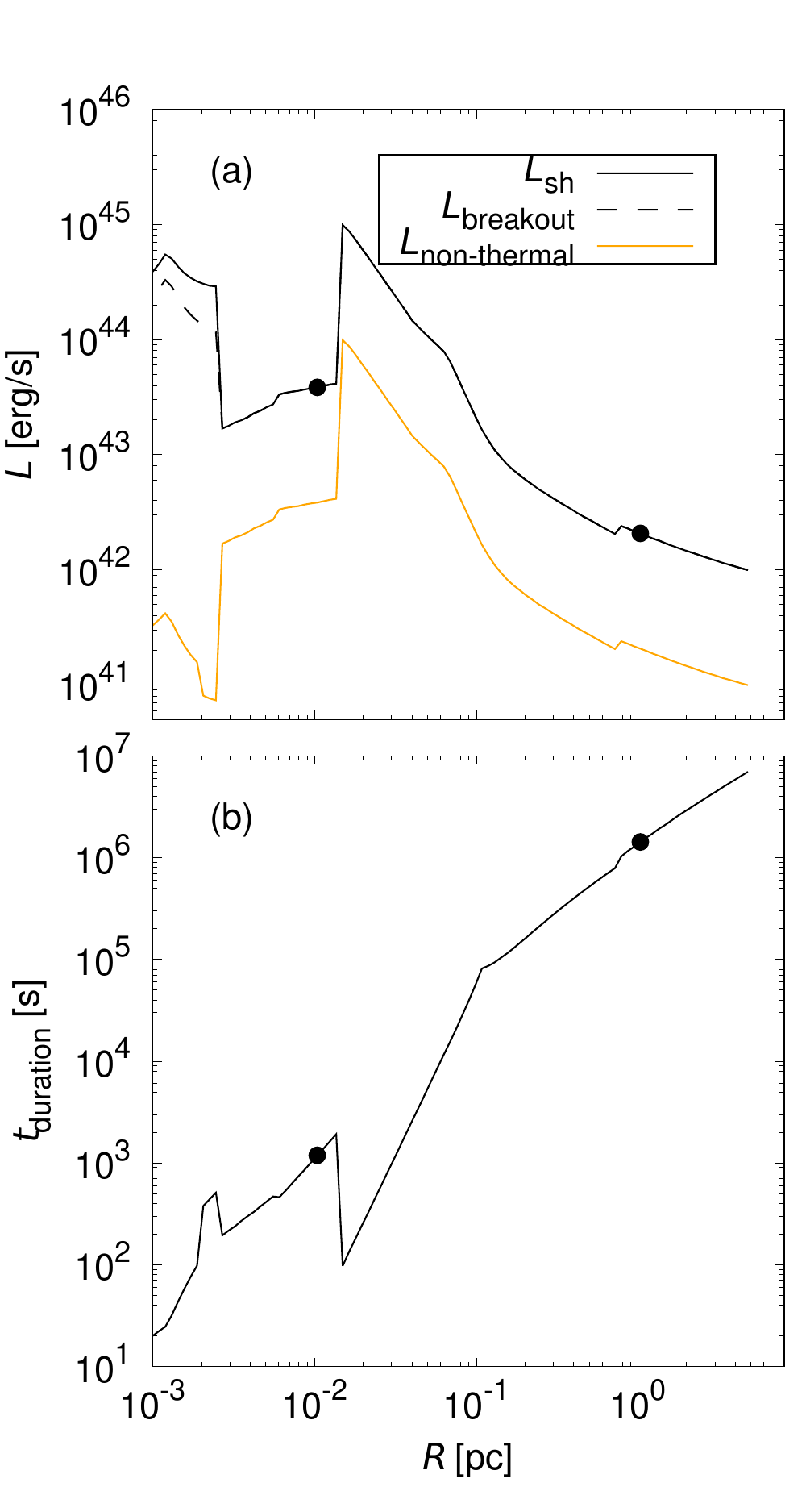}
\caption{
The luminosity and duration as a function of the distance from the SMBH ($R$) 
for emission from breakout of shocks produced around solitary BHs in the fiducial model. 
(a)~
The shock kinetic (solid black), breakout (dashed black), and non-thermal (solid orange) luminosity. 
(b)~
The duration of emission, $t_{\rm duration}$. 
The BH locations adopted in the fiducial model are indicated with filled circles superposed on the black solid lines. 
}
\label{fig:prop_fid_m6}
\end{center}
\end{figure}

\begin{figure}
\begin{center}
\includegraphics[width=90mm]{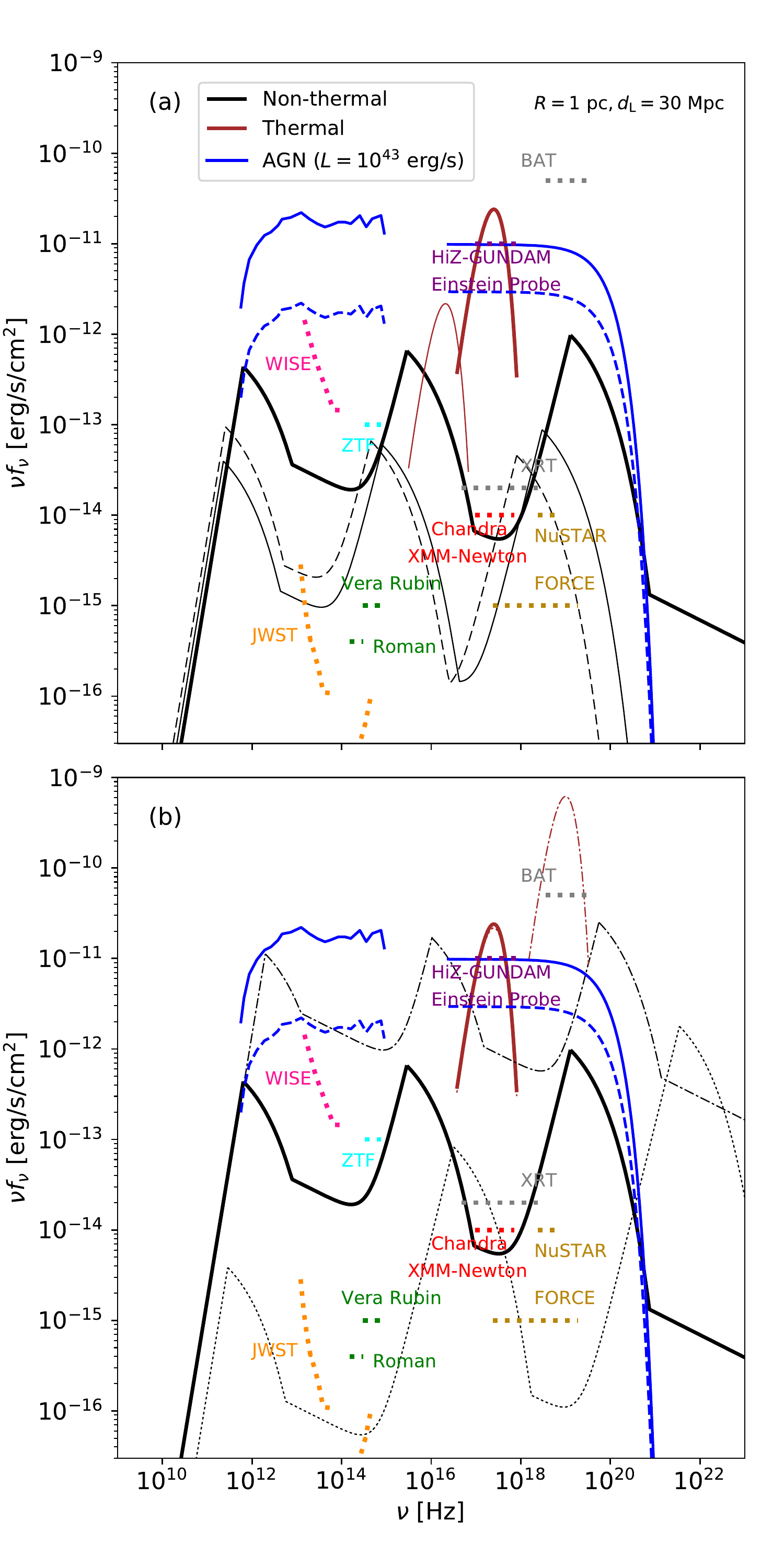}
\caption{
The spectral energy distribution for non-thermal (thick solid black) and thermal (thick solid brown) emission in the fiducial model (\S~\ref{sec:numerical_choice_m6}) at $R=1~{\rm pc}$. The left, middle, and right components in black lines represent synchrotron emission, synchrotron-self Compton (SSC), and second-order inverse Compton (IC) scattering, respectively. 
Blue and dashed blue lines represent emission from the host AGN and its variability, respectively. 
Dotted cyan, green, red, purple, gray, gold, pink, and orange lines indicate
the sensitivities of ZTF, Vera Rubin and Roman space telescope, Chandra and XMM-Newton, HiZ-GUNDAM and Einstein Probe, BAT and XRT, NuSTAR and FORCE, WISE, and JWST, respectively. 
The results are also shown for models with lower accretion rate onto BH ($f_{\rm acc}=0.1$, thin solid lines),
or lower efficiencies of 
electron acceleration ($\epsilon_{\rm e}=0.01$, thin dashed)
in panel~(a), 
or lower 
magnetic field amplification ($\epsilon_{\rm B}=10^{-5}$, thin dotted), 
or a higher jet efficiency ($\eta_{\rm j}=1$, thin dashed-dotted) in panel~b. 
}
\label{fig:l_nu1_m6}
\end{center}
\end{figure}

\begin{figure}
\begin{center}
\includegraphics[width=90mm]{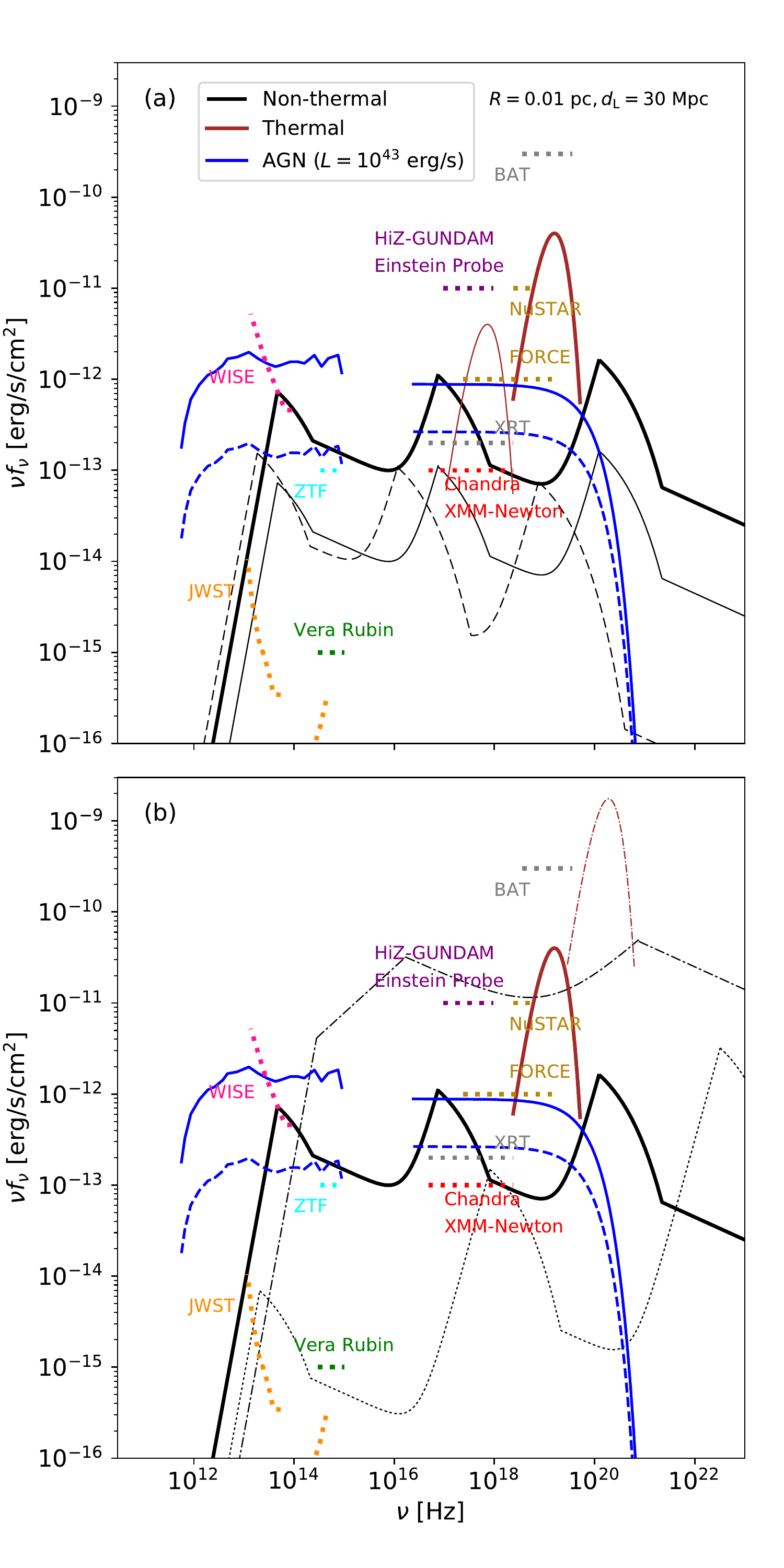}
\caption{
Same as Fig.~\ref{fig:l_nu1_m6}, but for BH sources at $R=0.01~{\rm pc}$. 
}
\label{fig:l_nu1_m6_001}
\end{center}
\end{figure}

\section{Properties of breakout emission}
\label{sec:properties_breakout}

In the following we discuss the 
properties of the breakout emission (see Paper~II for computational methods).

\subsection{Properties of breakout emission from solitary BHs}

\label{sec:properties_non_merging} 

 In the outer regions of $R\gtrsim 0.1~{\rm pc}$ for the fiducial model, since the aspect ratio ($H_{\rm AGN}/R$) of the AGN disk is large due to intense star formation to stabilize the disk \citep{Thompson05}, the accretion rates onto BHs, and accordingly the breakout luminosity $L_{\rm breakout}$, are low (Fig.\ref{fig:prop_fid_m6}~a). 
In the inner regions of $R\lesssim 10^{-2}~{\rm pc}$, 
$L_{\rm breakout}$ is low since gaps form in these regions, which reduce the accretion rates onto BHs. 
The accretion rates onto BHs at $R=1$ and $10^{-2}~{\rm pc}$ are, respectively,  
$3\times 10^{-4}~\Msun~{\rm yr^{-1}}$ and $3\times 10^{-3}~\Msun~{\rm yr^{-1}}$, corresponding to $\sim 10^4$ and $10^5$ times the Eddington rate. 
Despite the fact that the accretion rate and the duration of accretion are, respectively, much lower and longer than for gamma-ray bursts, 
the range of values for the plasma parameters ($\epsilon_{\rm e}$,  $\epsilon_{\rm B}$) can be reasonably adopted from observations of afterglow emission of gamma-ray bursts \citep[e.g.][]{Panaitescu2001,Santana2014}. 
This is motivated by the consideration that the basic physics of the radiation processes in the two contexts is similar, since in both cases the emission is produced by the interaction of a relativistic jet with the surrounding medium. 
However, we do note that the jet production physics may be different \citep[e.g.][]{Bromberg2016,Liu2017}, the compositions of the jets may be different \citep[e.g.][]{Beloborodov2003,Kimura2022,Chen2022}, 
the duration and the accretion rate are different, 
and the contribution of synchrotron self-absorption to emission is different due to the difference in the density of the ambient material. 
The above differences may result in significantly different distributions of the plasma parameters between the jets from BHs in AGN disks and those in gamma-ray bursts.

It is predicted that 
BHs tend to reside in the outer regions with $R\gtrsim {\rm pc}$ since the Type~I migration timescale is long, 
as well as in the inner regions with $R\lesssim 10^{-2}~{\rm pc}$ as annular gaps are predicted to form where migration is slow causing BHs to accumulate \citep[e.g.][]{Tagawa19,Gilbaum2021,Perna2021_AICs}. 
Therefore, below we consider the two cases with BHs at $1~{\rm pc}$ and at $10^{-2}~{\rm pc}$ as representative examples.

The spectral energy distributions for emission from BHs at $R=1~{\rm pc}$ and $R=0.01~{\rm pc}$ in the fiducial model are shown in Figs.~\ref{fig:l_nu1_m6} and ~\ref{fig:l_nu1_m6_001}, respectively. 
In this model, the thermal emission is computed under the assumption that the radiation energy in the shock is released once the shock becomes optically thin \citep[e.g.][]{Levinson2020}. Non-thermal emission is produced from electrons accelerated at collisionless shocks by 
synchrotron radiation, 
synchrotron-self Compton scattering, 
and second-order inverse Compton scattering.

The non-thermal emission is bright from the
infrared to the gamma-ray bands (solid black lines). 
The three peaks are contributed by synchrotron radiation, synchrotron-self Compton scattering, and second-order inverse Compton scattering, while the lower cutoff in non-thermal emission is due to synchrotron self-absorption. 
The thermal emission is mostly bright in X-rays (solid brown lines), and it has a much higher luminosity than the non-thermal emission (due to the reduction by a factor of $\epsilon_{\rm e}$ compared to the total energy).

In Paper~I 
we evaluated the breakout of shocks around a jet produced from a solitary BH and found it to be episodic. 
Emission phases last for the consumption timescale of the circum-BH disk with $t_{\rm cons}\sim 300~{\rm yr}$, followed by quiescent phases lasting for the resupply timescale of 
$t_{\rm re}\sim 10^4~{\rm yr}$ at $R\sim {\rm pc}$, with the cycle repeating. 
At $R\sim 10^{-2}~{\rm pc}$, 
$t_{\rm cons}\sim 1~{\rm yr}$ and $t_{\rm re}\sim 10~{\rm yr}$. 
By using the duration of emission (solid black line in Fig.~\ref{fig:prop_fid_m6}~b), 
which is $t_{\rm duration}\sim 2\times 10^6~{\rm s}$ at $R\sim {\rm pc}$ and $\sim 200~{\rm s}$ at $R\sim 0.01~{\rm pc}$, 
the total duration for breakout emission to be released from one BH over the AGN lifetime time is estimated to be $f_{\rm active}\sim t_{\rm duration}/t_{\rm re}\sim 10^{-5}$--$10^{-6}$. 
Here, note that the duration is reduced for the model with $\eta_{\rm j}=1$ to $7\times 10^5~{\rm s}$ at $R=1~{\rm pc}$ and $300~{\rm s}$ at $R=0.01~{\rm pc}$, 
and enhanced for the model with $f_{\rm acc}=0.1$ 
to $3\times 10^6~{\rm s}$ at $R=1~{\rm pc}$ and $6\times 10^3~{\rm s}$ at $R=0.01~{\rm pc}$. 
As discussed in Paper~I, 
the number of AGN disk-embedded BHs is $\sim 300~({\dot M}_{\rm in}/1~L_{\rm Edd}c^2)^{1/2}$. 
Using the predicted mass distribution of $dN_{\rm BH}/dR\propto R^{\gamma_\rho}$ with $-0.5 \lsim \gamma_\rho\lsim 0$ \citep{Freitag06,Hopman06,Alexander07}, 
we tentatively assume that $N_{\rm BH}\sim 200$ and $\sim 30$ BHs are embedded at $R\sim {\rm pc}$ and $\sim 10^{-2}~{\rm pc}$, respectively. 
Note that these numbers evolve with time and are highly uncertain. 
With these, the time interval between flares in one AGN is 
$t_{\rm interval}\sim t_{\rm re}/N_{\rm BH}\sim 50~{\rm yr}$ and 
$\sim 0.3~{\rm yr}$ at $R\sim {\rm pc}$ and $R\sim 10^{-2}~{\rm pc}$, respectively. 
From solitary BHs at $R\sim {\rm pc}$, 
thermal emission with luminosity of $\sim 2\times 10^{42}~{\rm erg/s}$ in X-ray bands (brown line in Fig.~\ref{fig:l_nu1_m6}) and non-thermal emission with $\sim 10^{39}$--$10^{41}~{\rm erg/s}$ in infrared to gamma-ray bands (black line in Fig.~\ref{fig:l_nu1_m6}) 
are predicted 
with duration of $t_{\rm duration}\sim 0.1~{\rm yr}$ (solid black line in Fig.~\ref{fig:prop_fid_m6}~b) by observing one AGN for $t_{\rm interval}\sim 50~{\rm yr}$. 
From solitary BHs at $R\sim 10^{-2}~{\rm pc}$, 
thermal emission with luminosity of $\sim 4\times 10^{43}~{\rm erg/s}$ in hard X-ray bands (brown line in Fig.~\ref{fig:l_nu1_m6_001}) 
and non-thermal emission with luminosity of $\sim 10^{41}$--$10^{42}~{\rm erg/s}$ in optical to gamma-ray bands 
(black line in Fig.~\ref{fig:l_nu1_m6_001}) 
are predicted 
with duration of $\lesssim 10^3~{\rm s}$ by observing an AGN for $\sim 0.3~{\rm yr}$.

Here, the luminosity from the host AGN in the relevant energy range is 
\begin{align}
\nu L_{\rm AGN}(\nu)\sim 10^{42}~{\rm erg/s}(M/10^6~\Msun)\nonumber\\
({\dot M} c^2/1~L_{\rm Edd})(f_{\rm bol}/10)^{-1}\,.
\end{align} 
where $f_{\rm bol}$ is the 
bolometric correction at the given frequency.
As depicted by the blue lines in Figs.~\ref{fig:l_nu1_m6} and \ref{fig:l_nu1_m6_001}, 
we assume that 
$f_{\rm bol}\sim 5$ at $c/\nu=4400~$\AA~ and extrapolate the luminosity 
for $10^{12}~{\rm Hz}\lesssim \nu \lesssim 10^{15}~{\rm Hz}$ using the cyan or blue points in Fig.~7 of \citet{Ho2008} depending on the assumed Eddington rate, 
and $f_{\rm bol}\sim 10$ in $0.1~{\rm keV}\leq h\nu$ \citep{Ho2008,Trakhtenbrot2017,Duras2020} with the upper exponential cut off at $300~{\rm keV}$ \citep[e.g.][]{Ricci2018}. 
We also assume that 
the fraction of the variable luminosity compared to the average luminosity ($f_{\rm var}$)
in optical bands with $t_{\rm duration}\lesssim 0.1~{\rm yr}$ is $f_{\rm var}\lesssim 0.1$ \citep{Kozlowski2016} 
and that in X-ray bands is $f_{\rm var}\sim 0.3$ (\citealt{Soldi2014,Maughan2019}, dashed blue lines).

In the optical, X-ray, and gamma-ray bands, the luminosity for non-thermal emission at $R=0.01~{\rm pc}$ exceeds the variable luminosity (solid black and dashed blue lines in Fig.~\ref{fig:l_nu1_m6_001}). 
Additionally, the variability of AGNs is typically stronger at shorter wavelengths \citep{Arevalo2008}, while non-thermal emission for $R=0.01~{\rm pc}$ is brighter at longer wavelengths in the optical bands. 
This unusual trend can help distinguish the breakout emission from a solitary BH from stochastic AGN variability. 
Also, the thermal X-ray luminosity clearly exceeds the AGN luminosity at both $R=1~{\rm pc}$ and $0.01~{\rm pc}$ (blue and brown lines in Figs.~\ref{fig:l_nu1_m6} and \ref{fig:l_nu1_m6_001}).
Hence, non-thermal emission from BHs at $R=0.01~{\rm pc}$ and thermal emission at $R=0.01~{\rm pc}$ and $R=1~{\rm pc}$ can be recognized as unusual variability of AGNs 
due to their luminosity and color.

However, we do note that the properties of emission are significantly influenced by uncertainties in the model parameters, namely $f_{\rm acc}$, $\epsilon_{\rm e}$, $\epsilon_{\rm B}$, and $\eta_{\rm j}$. 
Since $\epsilon_{\rm e}$ and $\epsilon_{\rm B}$ are respectively found to vary
within the ranges $\sim 0.01$--$0.3$ and $\sim 10^{-5}$--$0.1$ from GRB afterglow 
observations \citep{Panaitescu2001,Santana2014}, and $\eta_{\rm j}$ can typically range between $0.1$--$1$ as discussed above, in 
Figs.~\ref{fig:l_nu1_m6} and \ref{fig:l_nu1_m6_001} we also show emission models with $\epsilon_{\rm e}=0.01$ (thin dashed lines in panel~a), $\epsilon_{\rm B}=10^{-5}$ (thin dotted lines in panel~b), and $\eta_{\rm j}=1$ (thin dashed-dotted lines in panel~b). Given that $f_{\rm acc}$ is also highly uncertain, we additionally present a model with $f_{\rm acc}=0.1$ (thin solid lines in panel~a) as a representative example. 
Figs.~\ref{fig:l_nu1_m6} and \ref{fig:l_nu1_m6_001} show that non-thermal emission at $R=0.01~{\rm pc}$ can be dimmer than the AGN variability if the values of the model parameters are in their lower range. 
On the other hand, thermal emission is relatively less affected by these parameters. 
If we assume that $\eta_{\rm j}$ is well constrained by numerical simulations \citep[e.g.][]{Narayan2021}, then
thermal emission is almost solely influenced by the accretion rate onto the BHs. Thus, by observing the thermal emission, we can improve our understanding of the accretion processes in super-Eddington regimes.

As an additional point in relation to observations, 
we note that 
the estimates above 
suggest that we need to wait a long time to come across breakout emission by monitoring a single AGN. 
A more viable strategy would be that of observing many AGNs and check whether there is variability in various bands, as discussed in $\S~\ref{sec:observability_solitary}$. 

\subsection{Differences with respect to the emission from merging remnants}
\label{sec:difference}

While the basic physical emission mechanisms are the same for solitary BHs
and the post-merger BHs discussed in Paper~II, there are some important quantitative differences 
between the two cases, which we highlight below.
(1) Merger remnants tend to be massive compared to isolated BHs. 
(2) Merger remnants have higher BH spin magnitude, and hence higher conversion efficiency $\eta_{\rm j}$ of mass to jet power. 
(3) An enhancement of the accretion rate (compared to the solitary BH case) is expected for merger remnants due to shocks emerging in circum-BH disks by GW recoil kicks. 
(4) The flares from merger remnants are correlated with GW events. 
Due to (1)--(3), the luminosity of the breakout emission is higher in the case of a post-merger BH, and due to (4) the transients are easier to discover when produced by merger remnants and the associated GW source has already been detected. 
The above suggests that the emission from solitary BHs is more difficult to observe compared to that from merger remnants;  therefore, the search for emission from solitary BHs needs to be strategized, as discussed in the next section.

\subsection{Observability of breakout emission}
\label{sec:observability_solitary}

Here, we consider whether emission from solitary 
BHs can be discovered by current and future observing facilities.

The luminosity from non-thermal emission from solitary BHs at $R=0.01~{\rm pc}$ 
exceeds the sensitivity limit by ZTF, Vera Rubin, XRT, Chandra, XMM-Newton, WISE, and JWST 
at $d_{\rm L}=30~{\rm Mpc}$ (solid black, dashed cyan, dashed green, dashed gray, dashed red, and dashed pink, and dashed orange lines in Fig.~\ref{fig:l_nu1_m6_001}). 
Here, the typical variable luminosity of AGNs with duration of $\lesssim 0.1~{\rm yr}$ is \citep{Kozlowski2016}
\begin{align}
 \nu L_{\rm AGN,vari}(\nu)\sim 2\times 10^{40}~{\rm erg/s}(M/10^6~\Msun)\nonumber\\
 ({\dot M}c^2/0.01~L_{\rm Edd})(f_{\rm bol}/5)^{-1}(f_{\rm var}/0.1),
\end{align} 
which is generally lower than the luminosity of breakout emission $L_{\rm flare}\sim f_{\rm acc} (10^{41}$--$10^{42})~{\rm erg/s}$, unless the reduction in the accretion rate from the Bondi-Hoyle-Lyttleton rate is 
$f_{\rm acc}\lsim 0.1$. 
Additionally, non-thermal emission for $R=0.01~{\rm pc}$ is redder around the optical bands as mentioned in $\S\,$\ref{sec:properties_non_merging}. 
Thus, we can identify the breakout emission by the magnitude of its luminosity as well as by the
color of the flare. 
Here, note that the luminosity of the breakout emission is roughly proportional to the accretion rate onto the SMBH (e.g. Paper~I). Hence, it reduces the influence of both the AGN accretion rate ${\dot M}_{\rm in}$ and mass $M$ on the detectability of the breakout emission, since the AGN luminosity is also proportional to the accretion rate onto the SMBH. 
However, in pessimistic cases in which the accretion rate onto BHs is lower than assumed in the fiducial model, or the efficiencies of electron acceleration 
or magnetic field amplification are lower, then
non-thermal emission is dimmer than the typical variability of AGNs, and hence it would be difficult to observe. 

On the other hand, thermal emission from solitary BHs at $R=0.01~{\rm pc}$ and $R=1~{\rm pc}$ 
can most likely be discovered by 
several X-ray telescopes 
(Figs.~\ref{fig:l_nu1_m6} and \ref{fig:l_nu1_m6_001}), 
unless the accretion rate onto BHs is significantly lower than assumed in our fiducial model. 
For emission from $R=1~{\rm pc}$, 
since the flare is as rare as $t_{\rm interval}\sim 50~{\rm yr}$ per AGN, many ($\sim t_{\rm interval}/t_{\rm obs}\sim 50(t_{\rm interval}/50~{\rm yr})(t_{\rm obs}/{\rm yr})^{-1}$) AGNs 
need to be simultaneously observed to be discovered within the observational timescale ($t_{\rm obs}$). 
This requires wide field surveys, such as HiZ-GUNDAM/Einstein Probe (Table~\ref{table:telescope}). 
Multi-wavelength observations are likely to be key to identifying the breakout emission. This is because 
in the solitary BH model, flares occur simultaneously over a broad range of bands (infrared, optical, and X-ray), while for AGN variability the evolution is delayed depending on the frequency. 
The actual false-alarm probability for the detection of the breakout emission should be quantified in future work, using observed multi-band AGN light-curves.

We also estimate observability when the BH spin magnitude is maximal ($a_{\rm BH}\sim 1$), and so is the jet energy conversion efficiency  ($\eta_{\rm j} \sim 1$, \citealt{Narayan2021}), 
as an upper limit (dotted lines in panels~b of Figs.~\ref{fig:l_nu1_m6} and \ref{fig:l_nu1_m6_001}). 
In this case the luminosity of the breakout emission is enhanced by an order of magnitude compared to that in the fiducial model. Additionally, the shock velocity is higher by a factor of $\sim 1.6$ \citep{Bromberg2011}, and the frequency of the emission is enhanced accordingly. 
Then, the 
non-thermal emission from the BH at $R=0.01~{\rm pc}$ would be detectable by ZTF and 
the HiZ-GUNDAM/Einstein Probe. 
Additionally, the non-thermal emission for $R=1~{\rm pc}$ can be identified by 
wide-field surveys in radio bands, such as the Atacama Cosmology Telescope \citep{Naess2021} and the Large Submillimeter Telescope \citep{Kawabe2016}. 
Therefore, in this optimistic case, the emission could be easily discovered by several instruments at various wavelengths.

More generally, in the fiducial model 
non-thermal emission 
from solitary BHs at $R\sim 10^{-2}~{\rm pc}$ can be discovered by ZTF, Vera Rubin, XRT, Chandra, XMM-Newton, JWST, and WISE as unusually intense and red-colored variability, 
while 
thermal emission from solitary BHs 
at $R\sim 1~{\rm pc}$ can be discovered by 
HiZ-GUNDAM and Einstein Probe.

\section{Conclusions}

In this paper
we have evaluated the properties of breakout emission from shocks emerging around jets launched from accreting and spinning solitary BHs embedded in AGN disks, and discussed the observability of such emission. 
In our model, accretion, and hence jet formation, is episodic, since gas around the BHs is evacuated by the jets; once gas is resupplied, the jet is expected to collide with the gas. Due to the formation of shocks at collision, thermal emission produced by the shocked gas and non-thermal emission produced by accelerated electrons are expected. 
Our main results are summarized as follows:

\begin{enumerate}

\item 
Thermal and non-thermal emission are bright in X-ray bands and in infrared to gamma-ray bands, respectively.

\item 
Breakout-emission from solitary BHs is harder to observe than from merger remnants because (1) the non-thermal and thermal emission are not as bright, and (2) the burst is rare 
and there is no GW trigger. Hence, catching it requires monitoring a large number of AGNs.  However, we can still identify breakout emission from solitary BHs as peculiar flares in nearby AGNs, characterized by broad-band non thermal emission (from the $\gamma$-rays to the IR), with superimposed thermal emission
and duration that depends on the distance of the BH from the central SMBH, varying between $10^3-10^6$~s for distances $R\sim 0.01-1$~pc.

\item 
Non-thermal 
emission from solitary BHs at $R~\sim~0.01~{\rm pc}$ from the SMBH with  duration of $\sim 10^3~{\rm s}$ can be discovered by infrared, optical, and X-ray telescopes 
as unusually red-colored variability of less luminous AGNs. 
Additionally, thermal emission from solitary BHs 
at $R\sim 1~{\rm pc}$ with  duration of $\sim 10^6~{\rm s}$ 
can be discovered by current and future X-ray telescopes. 

\end{enumerate}

We find that the observability of the breakout emission from solitary BHs in AGN disks is strongly influenced by accretion processes in super-Eddington regimes. 
To discover signatures from the solitary BHs, 
the accretion processes 
and plasma physics
should be better understood through numerical simulations.
Conversely, 
if the emission is discovered 
but their properties are different from what we predict, this would improve our understanding of the underlying accretion processes and plasma physics.

\acknowledgments

This work was financially supported 
by Japan Society for the Promotion of Science (JSPS) KAKENHI 
grant Number JP21J00794 (HT) and 22K14028 (S.S.K.). 
S.S.K. was supported by the Tohoku Initiative for Fostering Global Researchers for Interdisciplinary Sciences (TI-FRIS) of MEXT's Strategic Professional Development Program for Young Researchers.
Z.H. was supported by NASA grant NNX15AB19G and NSF grants AST-2006176 and AST-1715661.
R.P. acknowledges support by NSF award AST-2006839.
I.B. acknowledges the support of the Alfred P. Sloan Foundation and NSF grants PHY-1911796 and PHY-2110060.

\clearpage
\newpage

\appendix

\section*{Telescopes}

We list the name and properties of telescopes in Table~\ref{table:telescope}.

\begin{table}
\begin{center}
\caption{The name and properties of telescopes appropriate for detecting electromagnetic signatures from solitary BHs in AGN disks.}
\label{table:telescope}
\hspace{-5mm}
\begin{tabular}{c|c|c|c|c}
\hline 
Telescope name & Photon energy
& Sensitivity $[{\rm erg/s/cm^2}]$& Field of view$[{\rm sr}]$ & Operation\\
\hline\hline
JWST& $\sim 0.04$--$2~[{\rm eV}]$&$\sim 10^{-17}$--$10^{-15}$ for $t_{\rm int}\sim10^4~{\rm s}$&$\sim 10^{-6}$&pointing telescope\\\hline
WISE& $\sim 0.05$--$0.4~[{\rm eV}]$&$\sim 10^{-13}$--$10^{-12}$ for $t_{\rm int}\sim10^4~{\rm s}$&$\sim 10^{-4}$&pointing telescope\\\hline
Roman space telescope \citep{Spergel_2015}
& $\sim 0.6$--$1~[{\rm eV}]$&$\sim 4\times 10^{-16}$ for $t_{\rm int}\sim10^4~{\rm s}$&$\sim 0.6$&wide-field survey\\\hline
ZTF \citep{Bellm_2018}& 
$\sim 1.4$--$3.1~[{\rm eV}]$
&$\sim 10^{-13}$ for $t_{\rm int}\sim30~{\rm s}$
&0.01&wide-field survey\\\hline
Vera Rubin \citep{Ivezic2019}&
$\sim 1.2$--$3.9~[{\rm eV}]$
&$\sim 10^{-15}$ for $t_{\rm int}\sim40~{\rm s}$&0.003&wide-field survey\\\hline
Subaru/HSC \citep{Aihara2018}& 
$\sim 1$--$3~[{\rm eV}]$
&$\sim 10^{-16}$--$10^{-15}$ for $t_{\rm int}\sim10^3~{\rm s}$
&
0.0005&pointing telescope
\\\hline
Tomo-e Gozen \citep{Tomoe2018}& 
$\sim 1.7$--$3.4~[{\rm eV}]$
&$\sim 2\times 10^{-13}$ 
for 
$t_{\rm int}\sim 100~{\rm s}$
&0.006&wide-field survey
\\\hline
Chandra& $\sim 0.2$--$10~[{\rm keV}]$&$\sim10^{-14}$  for $t_{\rm int}\sim 2\times 10^4~{\rm s}$&$6\times 10^{-5}$&pointing telescope\\\hline
XMM-Newton \citep{Jansen2001}& $\sim 0.4$--$3~[{\rm keV}]$&$\sim10^{-14}$ for $t_{\rm int}\sim 10^4~{\rm s}$&$8\times 10^{-5}$&pointing telescope\\\hline
HiZ-GUNDAM \citep{Yonetoku2020c}& $\sim 0.4$--$4~[{\rm keV}]$&$\sim10^{-11}$ for $t_{\rm int}\sim10^4~{\rm s}$&1.2&wide-field survey\\\hline
Einstein Probe \citep{Yuan2015}& $\sim 0.5$--$4~[{\rm keV}]$&$\sim3\times 10^{-11}$ for $t_{\rm int}\sim 10^3~{\rm s}$&1.0&wide-field survey\\\hline
MAXI \citep{Matsuoka2009_MAXI}& $\sim 2$--$30~[{\rm keV}]$&$\sim7\times 10^{-11}$ for $t_{\rm int}\sim 6\times 10^5~{\rm s}$&0.07&wide-field survey\\\hline
NuSTAR \citep{Harrison2013}
& $\sim 10$--$30~[{\rm keV}]$&$\sim 10^{-14}$ for $t_{\rm int}\sim 10^6~{\rm s}$&$3\times10^{-5}$&pointing telescope
\\\hline
FORCE \citep{Mori16_FORCE}& $\sim 1$--$80~[{\rm keV}]$&$\sim 10^{-14}(t_{\rm int}/10^5~{\rm s})^{-1}$&$10^{-5}$&pointing telescope\\\hline
{\it Swift} X-ray telescope (XRT) \citep{Burrows2005}& $\sim 0.2$--$10~[{\rm keV}]$&$\sim 2\times 10^{-14}(t_{\rm int}/10^{4}~{\rm s})^{-1}$&$5\times 10^{-5}$&pointing telescope\\\hline
{\it Swift} BAT \citep{Barthelmy2005}& $\sim 15$--$150~[{\rm keV}]$&$\sim 10^{-8}(t_{\rm int}/1~{\rm s})^{-1/2}$&1.4&wide-field survey\\\hline
{\it Fermi} GBM \citep{Meegan2009}& $\sim 8$--$4000~[{\rm keV}]$&$\sim 10^{-8}$--$10^{-6}$ for $t_{\rm int}\sim1~{\rm s}$&$\sim 4\pi$&wide-field survey\\\hline
{\it INTEGRAL} SPI-ACS \citep{Winkler2003}& $\sim 75$--$2000~[{\rm keV}]$&$\sim 10^{-7}$--$10^{-6}$ for $t_{\rm int}\sim1~{\rm s}$&$\sim 4\pi$&wide-field survey\\\hline
\end{tabular}
\end{center}
\end{table}

\clearpage
\newpage

\bibliographystyle{aasjournal}
\bibliography{agn_bhm}


\end{document}